\documentclass[twocolumn]{jpsj2}

\newcommand{\bsigma}{\mbox{\boldmath $\sigma$}}

\usepackage{amsmath}
\usepackage{graphicx}

\author{
Ken-ichi \textsc{Sasaki} \thanks{E-mail: sasaken@flex.phys.tohoku.ac.jp}
and Riichiro \textsc{Saito}
}

\inst{
Department of Physics, Tohoku University and CREST, JST,
Sendai 980-8578, Japan
}
\recdate{\today}

\title{Magnetism as a mass term of the edge states in graphene}

\abst{
 The magnetism by the edge states in graphene 
 is investigated theoretically.
 An instability of the pseudo-spin order of the edge states
 induces ferrimagnetic order in the presence of the
 Coulomb interaction.
 Although the next nearest-neighbor hopping 
 can stabilize the pseudo-spin order, 
 a strong Coulomb interaction makes the pseudo-spin unpolarized 
 and real spin polarized.
 The magnetism of the edge states makes two peaks 
 of the density of states
 in the conduction and valence energy bands near the Fermi point.
 Using a continuous model of the Weyl equation,
 we show that the edge-induced gauge field
 and the spin dependent mass terms are 
 keys to make the magnetism of the edge states.
 A relationship between the magnetism of the edge states 
 and the parity anomaly is discussed.
}

\begin{document}

\maketitle

\section{introduction}

The electronic properties of graphene~\cite{novoselov05,zhang05,heersche07}
have attracted much attention
mainly because of its relativistic character of
low energy electronic excitation.
The energy band structure of graphene
consists of two Dirac cones at the K and K' points
in the $k$-space.
The electron dynamics around each Dirac point
is approximated by the Weyl equation which
describes a massless particle.
When we consider solutions of the Weyl equation
for a finite (or semi-finite) graphene cluster (ribbon)
with the zigzag edges, the spatially-localized
edge states~\cite{fujita96} exist
around the Fermi energy.~\cite{sasaki06jpsj} 
The existence of the edge states 
depends on the shape of the edge
for a graphene cluster.
For example, the zigzag edge yields the edge
states while the armchair edge does not.
The energy dispersion for the edge states which
appears only between the two Dirac points 
smoothly connects to the energy dispersion of the
delocalized states.~\cite{sasaki08ab}
A large local density of states (LDOS)
by the edge states may induce
magnetism~\cite{fujita96} and
superconductivity~\cite{sasaki06super} 
near the zigzag edge.
The existence of a mass for the Dirac particle
is an important issue because 
it gives rise to an energy gap at the Dirac points
and relates to the ordered states.
We examine a mechanism that the Coulomb interaction
makes a mass and a magnetism.

Fujita {\it et al}. discussed 
the effect of the Coulomb interaction on the edge state, 
in which the electronic spins are localized at the edge 
to form a ferromagnetic state at one zigzag edge
and another ferromagnetic state with the opposite
spin at another zigzag edge.~\cite{fujita96} 
The occurrence of the magnetism is investigated 
by first principles calculations, too.~\cite{okada03,kusakabe03,son06} 
If the ferromagnetic state appeared at one edge, 
we would expect that two peaks for up and down spin states
appear in scanning tunneling spectroscopy (STS).
However this situation seems to be inconsistent with 
the STS measurements 
in which they observed only one LDOS peak
near the zigzag edge of
graphite.~\cite{klusek00,kobayashi05,kobayashi06,niimi05,niimi06}  
Klusek {\it et al.}~\cite{klusek00} 
found a peak of LDOS in the energy range of 20-250 meV 
above the Fermi level at the edges of circular pits 
on graphite surface.
Kobayashi {\it et al.}~\cite{kobayashi05,kobayashi06} 
and Niimi {\it et al.}~\cite{niimi05,niimi06} independently
observed a peak in the LDOS
below the Fermi energy by 20 $\sim$ 30 meV.
Since the peak appears only conduction (Klusek {\it et al.}) 
or valence energy band (Kobayashi {\it et al.} and Niimi {\it et al.}),
it suggests that the edge states do not make a magnetism.
Thus it is an interesting problem for understanding 
the occurrence of the ferromagnetic order at the edge 
in the presence of the Coulomb interaction. 
When we see the calculated results of Fujita {\it et al}.,
the polarized spin appears for the edge states at a much small
on-site Coulomb interaction $U$ value compared 
with the nearest neighbor interaction $\gamma_0$
(see Fig.~5 in Ref.~\citen{fujita96}). 
Although we reproduce their results numerically, 
the results are very surprising. 
A possible reason why the spin ordering occurs 
for such a small U 
is due to a special fact that the wavefunction of the edge
states has an amplitude only one of the two
sublattices (A and B) and thus 
the nearest neighbor interaction is suppressed. 
When we introduce the next nearest neighbor interaction, $\gamma_n$, 
the spin polarization around $U/\gamma_0\sim 0$ disappears 
and spin ordering appears from finite values of $U$
depending on $\gamma_n$, which we will show in this paper.

The wavefunction for two sublattice structure 
is referred to as the ``pseudo-spin''.
An edge state can be described by a pseudo-spin polarized
state.~\cite{sasaki08ab} 
A pseudo-spin structure gives a rich variety of interesting physical
phenomena not only the edge states but also the extended states. 
For example, the absence of backward scattering mechanism 
is relevant to this pseudo-spin nature,~\cite{ando98,ando05}
in which a $2\pi$ rotation of a pseudo-spin wavefunction
around the K-point in the two-dimensional Brillouin zone 
does not gives the original wavefunction 
but gives minus sign to the wavefunction.
Thus, the pseudo-spin is quite similar to the real spin
in the real space.
In this paper, we show that the pseudo-spin also plays an important role
for the magnetism (or real spin) of the edge states, 
which is shown by a numerical analysis of the lattice model
and by a analytical study of the Weyl equation.

This paper is organized as follows.
In Sec.~\ref{p-spin} we explain the model Hamiltonian
and introduce symmetric and antisymmetric variables for the pseudo-spin.
In Sec.~\ref{sec:ana}
we show numerical results for the ground state of the model.
In Sec.~\ref{sec:cm} we use a continuous model to 
examine the mechanism of the magnetism of the edge states.
Discussion and summary are given in Sec.~\ref{sec:dis}.

\section{pseudo-spin representation of Hamiltonian}\label{p-spin}

The Hamiltonian is given by
${\cal H}_{0\uparrow}+{\cal H}_{0\downarrow}+{\cal H}_U$ where
${\cal H}_{0s}\equiv -\gamma_0 
\sum_{\langle i,j \rangle} c_{s,i}^\dagger c_{s,j}$
($s=\uparrow,\downarrow$)
is the nearest-neighbor tight-binding Hamiltonian
($\gamma_0\approx 3$eV is the hopping integral),
and ${\cal H}_U$ is the Hubbard on-site interaction.
${\cal H}_U$ is written as 
\begin{align}
 {\cal H}_U = 
 U \sum_{\bf r} 
 n_\uparrow({\bf r}) n_\downarrow({\bf r}),
 \label{eq:H_U}
\end{align}
where $U$ is the on-site energy and
$n_\uparrow({\bf r})$ ($n_\downarrow({\bf r})$)
is the density operator 
of up (down) spin electron at site ${\bf r}$.
Since the hexagonal lattice consists of 
two sublattice, {\rm A} and {\rm B}, 
${\cal H}_U$ is given as a summation over unit cells as
\begin{align}
 {\cal H}_U = U \sum_{{\bf r}_u}
 \sum_{p={\rm A},{\rm B}}
 n_{\uparrow,p}({\bf r}_u) n_{\downarrow,p}({\bf r}_u),
 \label{eq:H_U_unit}
\end{align}
where $n_{\uparrow,p}({\bf r}_u)$ ($n_{\downarrow,p}({\bf r}_u)$) 
is the density operator of up (down) spin electron at $p$-sublattice
($p= {\rm A},{\rm B}$), and 
${\bf r}_u$ denotes the position of a unit cell.
For a unit cell, 
we introduce a density and a magnetization at ${\bf r}_u$ as
$n_{p}({\bf r}_u)=n_{\uparrow,p}({\bf r}_u)+n_{\downarrow,p}({\bf
r}_u)$ and $m_{p}({\bf r}_u)=n_{\uparrow,p}({\bf
r}_u)-n_{\downarrow,p}({\bf r}_u)$, respectively.
Hereafter ${\bf r}_u$ for each variable is omitted for simplicity.
From $n_{p}$ and $m_{p}$,
we define density and magnetization 
for a unit cell as
\begin{align}
 n = n_{\rm A} + n_{\rm B}, \ \
 m = m_{\rm A} + m_{\rm B}.
 \label{eq:nm}
\end{align}
$n$ and $m$ are symmetric with respect to the sublattice.
Here we introduce pseudospin order and antiferromagnetic order
for a unit cell,
\begin{align}
 p_n=n_{\rm A}-n_{\rm B}, \ \
 p_m=m_{\rm A}-m_{\rm B},
 \label{eq:pnm}
\end{align}
which are anti-symmetric with respect to the sublattice.
$p_n$ ($p_m$) represents charge (spin) polarization
within the hexagonal unit cell.
${\cal H}_U$ can be rewritten in terms of $n,m,p_n$ and $p_m$ as
\begin{align}
 {\cal H}_U = \frac{U}{8} \sum_{{\bf r}_u}
 \left( n^2 + p_n^2 - m^2 - p_m^2 \right).
 \label{eq:HU}
\end{align}
This representation of ${\cal H}_U$ shows that 
not only non-vanishing magnetization 
($\langle m \rangle \ne 0$) 
but also antiferromagnetic 
($\langle m \rangle=0$ and $\langle p_m \rangle\ne 0$) 
or ferrimagnetic
($\langle m \rangle\ne 0$ and $\langle p_m \rangle\ne 0$) 
spin configuration 
are favored to decrease ${\cal H}_U$ where
$\langle {\cal O} \rangle$ denotes 
the expectation value of operator ${\cal O}$ 
for the ground state.

By applying the mean-field approximation to Eq.~(\ref{eq:H_U_unit}),
${\cal H}_U = U \sum_{{\bf r}_u,p} 
 \langle n_{\uparrow,p} \rangle n_{\downarrow,p} 
+n_{\uparrow,p} \langle n_{\downarrow,p} \rangle 
-\langle n_{\uparrow,p} \rangle \langle n_{\downarrow,p} \rangle$,
the Hamiltonians for up and down spin electrons 
in graphene are given as follows:
\begin{align}
 \begin{split}
  & {\cal H}_{\uparrow} \equiv 
  {\cal H}_{0\uparrow} + \frac{U}{2}\sum_{{\bf r}_u}
  \begin{pmatrix}
   \langle n_{\rm A} - m_{\rm A} \rangle & 0 \cr 
   0 & \langle n_{\rm B} - m_{\rm B} \rangle
  \end{pmatrix}
  \begin{pmatrix}
   n_{\uparrow,{\rm A}} \cr n_{\uparrow,{\rm B}}
  \end{pmatrix},
  \\
  & {\cal H}_{\downarrow} \equiv
  {\cal H}_{0\downarrow} + \frac{U}{2}\sum_{{\bf r}_u}
  \begin{pmatrix}
   \langle n_{\rm A} + m_{\rm A} \rangle & 0 \cr 
   0 & \langle n_{\rm B} + m_{\rm B} \rangle
  \end{pmatrix}
  \begin{pmatrix}
   n_{\downarrow,{\rm A}} \cr n_{\downarrow,{\rm B}}
  \end{pmatrix}.
 \end{split}
 \label{eq:Hd}
\end{align}
The Hamiltonians of Eq.~(\ref{eq:Hd})
can also be rewritten 
using Eqs.~(\ref{eq:nm}) and (\ref{eq:pnm}) as
\begin{align}
 \begin{split}
  & {\cal H}_{\uparrow} = 
  {\cal H}_{0\uparrow} 
  + \frac{U}{4} 
  \sum_{{\bf r}_u} \left[
  \left(\langle p_n \rangle - \langle p_m \rangle \right) \sigma_z   
  + \left(\langle n \rangle - \langle m \rangle \right) I \right] 
  \begin{pmatrix}
   n_{\uparrow,{\rm A}} \cr n_{\uparrow,{\rm B}}
  \end{pmatrix},
  \\
  & {\cal H}_{\downarrow} = 
  {\cal H}_{0\downarrow} + 
  \frac{U}{4} \sum_{{\bf r}_u}
  \left[
  (\langle p_n \rangle + \langle p_m \rangle) \sigma_z
  + (\langle n \rangle +\langle m \rangle) I \right] 
  \begin{pmatrix}
   n_{\downarrow,{\rm A}} \cr n_{\downarrow,{\rm B}}
  \end{pmatrix},
 \end{split}
 \label{eq:Hspin}
\end{align}
where $\sigma_z={\rm diag}(1,-1)$ 
and $I={\rm diag}(1,1)$.
The pseudo-spin variables are proportional to $\sigma_z$
and affect magnetization of the edge states 
as we will show in the following sections.

\begin{table}[htbp]
 \caption{\label{tab:parity}Parities with respect to spin and pseudo-spin}
  \begin{tabular}{lllllll}
   \hline
   & $p_n$ & $p_m$ & $n$ & $m$ & $\sigma_z$ & Coupling\\
   \hline
   Spin ($\uparrow \leftrightarrow \downarrow$) 
   & $+$ & $-$ & $+$ & $-$ & $+$ & ${\bf B}^{\rm em}$
   \\
   Pseudo-Spin ({\rm A}$\leftrightarrow${\rm B}) 
   & $-$ & $-$ & $+$ & $+$ & $-$ & ${\bf B}^{\rm q}$\\
   \hline
  \end{tabular}
% \end{ruledtabular}
\end{table}
In Table.~\ref{tab:parity},
we show the parities of $p_n$, $p_m$, $n$, $m$ and $\sigma_z$
for changing the direction of spin and pseudo-spin.
%The parity with respect to spin manifests the relationship between 
%${\cal H}_\uparrow$ and ${\cal H}_\downarrow$ in Eq.~(\ref{eq:Hspin}).
The interaction terms in Eq.~(\ref{eq:Hspin})
are invariant with respect to pseudo-spin parity: 
$p_n\to -p_n$,
$p_m\to -p_m$, $n \to n$, $m \to m$
and $\sigma_z \to -\sigma_z$,
and we have 
${\cal H}_{\uparrow} \leftrightarrow {\cal H}_{\downarrow}$
for spin parity: $p_n\to p_n$,
$p_m\to -p_m$, $n \to n$, $m \to -m$
and $\sigma_z \to \sigma_z$.
%Altough the Hubbard model is a special model,
%the interaction term is most general from the point of view of 
%spin and pseudo-spin symmetries.
Since ${\cal H}_U$ is even parity with respect to spin and pseudo-spin, 
it is expected that 
a ground state is realized by spontaneous symmetry breaking
if ${\cal H}_0$ is symmetric, too.
${\cal H}_0$ is symmetric with respect to spin 
but asymmetric with respect to pseudo-spin 
due to the presence of the zigzag boundary.
This can be explained as follows.
A magnetic field, ${\bf B}^{\rm em}$, 
breaks the spin degeneracy of the ground state by
the Zeeman term, $-{\bf B}^{\rm em}\cdot {\bf m}$, 
which is odd parity with respect to spin.
Similarly, we can define a pseudo-magnetic field, 
${\bf B}^{\rm q}$,
that couples to $\sigma_z$ (see Eq.~(\ref{eq:second-H})) and
breaks the degeneracy of the pseudo-spin parity.
In fact, 
it can be shown that ${\bf B}^{\rm q}$ appears 
at the zigzag edge~\cite{sasaki06jpsj}
so that ${\cal H}_0$ can induce a pseudo-spin order, 
$\langle p_n \rangle \ne 0$, near the edge.
In the next section, 
we will show numerically that ${\cal H}_0$ of a zigzag nanotube 
breaks the pseudo-spin parity of $p_n$, which is an important 
to obtain magnetism for 
the total Hamiltonian,
${\cal H}_{\uparrow}+{\cal H}_{\downarrow}$.

\section{numerical results and analysis}\label{sec:ana}

In this section, 
we show numerical results for the ground state of 
${\cal H}_{\uparrow}+{\cal H}_{\downarrow}$.
$\langle n/4 \rangle$, $\langle p_n \rangle$, 
$\langle p_m \rangle$, and $\langle m \rangle$ are plotted 
for $(50,0)$ zigzag nanotube with length $L\approx 4$nm, and
LDOS curves are calculated for 
$(100,0)$ zigzag nanotube with length $L\approx 20$nm.
%The system size is large enough to be applicable to graphene.
We set the origin of the Fermi energy $E_{\rm F}=0$ as 
${\cal H}_s = U/2$.

\begin{figure}[htbp]
 \begin{center}
  \includegraphics[scale=0.45]{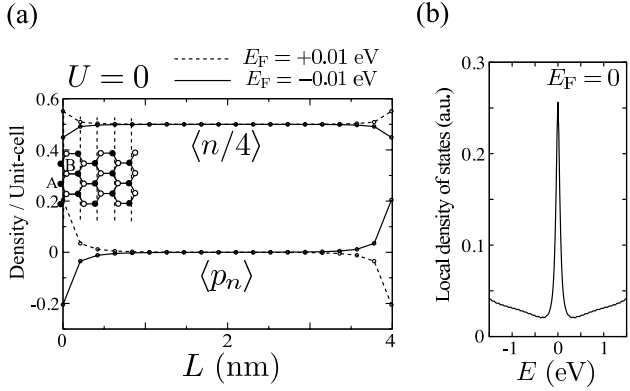}
 \end{center}
 \caption{
 (a) Self-consistent solution of $\langle n/4 \rangle$
 and $\langle p_n \rangle$ for $U=0$ and $E_{\rm F}=\pm 0.01$ eV.
 Due to the presence of the edge states consisting a flat energy band at
 $E_{\rm F}=0$, $\langle n/4 \rangle$ and $\langle p_n \rangle$
 are unstable against the small change of $E_{\rm F}$.
 $\langle {\cal O}({\bf r}_u) \rangle$ 
 depends only on the distance from an edge
 due to rotational symmetry around the axis of the tube.
 (b) Corresponding LDOS curve at $L\approx 0.5$ nm.
 The degenerate spin up and down edge states 
 make a sharp peak at $E_{\rm F}=0$ in the case of $U=0$.
 }
 \label{fig:n_pn}
\end{figure}

In Fig.~\ref{fig:n_pn}(a), 
we plot $\langle n/4 \rangle$ and $\langle p_n \rangle$ 
in the case of $U=0$.
Since $U=0$, 
$\langle m \rangle=\langle p_m \rangle=0$. 
The solid (dashed) curves 
are the results for the Fermi energy 
$E_{\rm F}=-0.01$ eV ($+0.01$ eV).
$\langle n/4 \rangle$ and $\langle p_n \rangle$ 
are modulated near the edges
and their difference from the constant values 
($\langle n/4 \rangle=0.5$ and $\langle p_n \rangle=0$) 
is due to the presence of the edge states.~\cite{fujita96} 
The wavefunction of the edge states is localized near the edges 
so that $\langle n/4 \rangle$ is different from 0.5 
(i.e., half filling $n=2$) only near the edges.
Moreover, the wavefunction of the edge states
is polarized about the pseudo-spin and 
$\langle p_n \rangle$ is nonzero, too.
$\langle n/4 \rangle$ and $\langle p_n \rangle$ are unstable 
against a small change of $E_{\rm F}$ due to 
the flat energy band of the edge states.
In Fig.~\ref{fig:n_pn}(b), 
we plot the LDOS at $L\approx 0.5$ nm
in the case of $E_{\rm F}=0$.
The spin up edge states and spin down edge states are degenerate
in the case of $U=0$ so that 
they make a sharp LDOS peak at $E_{\rm F}=0$.
To show the (smooth) LDOS curve, 
we put a constant width ($0.05$ eV) for each state.

%%%%%%%%%%%%%%%%%%%%%%%%%%%%%
%%%%%%%%%%%%%%%%%%%%%%%%%%%%%

\begin{figure}[htbp]
 \begin{center}
  \includegraphics[scale=0.45]{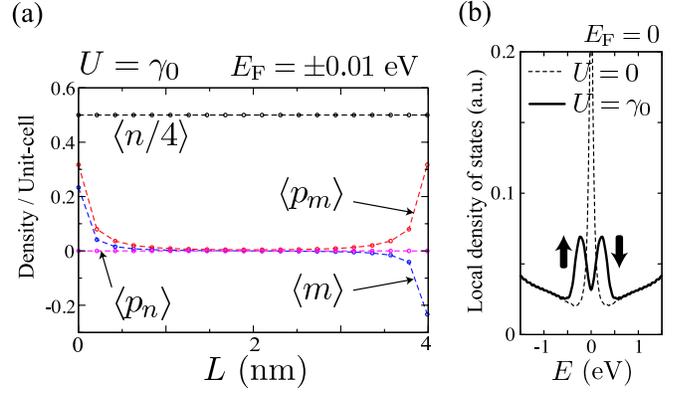}
 \end{center}
 \caption{(color online)
 (a) 
 Self-consistent solution of $\langle n/4 \rangle$, $\langle m \rangle$,
 $\langle p_n \rangle$ and $\langle p_m \rangle$ for $U=\gamma_0$ and 
 $E_{\rm F}=\pm 0.01$eV.
 The spin around the edge is polarized and 
 $|\langle p_m \rangle| \approx |\langle m \rangle|$
 shows that spin is a ferrimagnetic configuration.
 (b) The LDOS curve at $L\approx 0.5$ nm shows that
 spin up edge states (denoted by up-arrow) and 
 spin down edge states (denoted by down-arrow)
 make two sharp peaks (solid curve)
 near $E_{\rm F}=0$ when $U \ne 0$.
 }
 \label{fig:density_U}
\end{figure}

In Fig.~\ref{fig:density_U}(a), 
we plot self-consistent solution of
$\langle n/4 \rangle$, $\langle m \rangle$, $\langle p_n \rangle$,
and $\langle p_m \rangle$ 
in the case of $U=\gamma_0$.
The result shows that, 
$\langle n/4 \rangle \approx 0.5$ and 
$\langle p_n \rangle \approx 0$ 
hold at each hexagonal unit cell, and 
$\langle m \rangle$ and $\langle p_m \rangle$ 
become nonzero near the zigzag edges.
Since ${\cal H}_U$ can stabilize 
$\langle n \rangle$ and $\langle p_n \rangle$
according to Eq.~(\ref{eq:HU}),
the pseudo-spin polarization ($\langle p_n \rangle \ne 0$)
which exists for $U=0$ disappears for the ground state
due to a finite value of $U$.
The corresponding LDOS curve is shown in Fig.~\ref{fig:density_U}(b). 
Because of $U$,
the spin up (down) edge states 
are shifted above (below) the Fermi energy 
so that the ferrimagnetic order 
($\langle m \rangle\ne 0$ and $\langle p_m \rangle\ne 0$) 
appears and that 
two LDOS peaks appear around $E_{\rm F}=0$.
The LDOS curve in the case of $U=0$ is also shown 
in Fig.~\ref{fig:density_U}(b) for comparison.

%%%%%%%%%%%%%%%%%%%%%%%%%%%%%
%%%%%%%%%%%%%%%%%%%%%%%%%%%%%

The numerical results can be explained qualitatively using
Eq.~(\ref{eq:Hspin}) as follows.
When $\langle p_n \rangle=0$ and $\langle p_m \rangle > 0$,
the energy of a spin up electron is shifted below 
the Fermi energy for a pseudo-spin up state ($\sigma_z =1$).
On the other hand, 
the energy of a spin down electron is shifted above
the Fermi energy for the same state (a pseudo-spin up state).
Thus, when $E_{\rm F}= 0$, 
the ground state has a finite positive value of $\langle p_m \rangle$
and $\langle m \rangle$,
which lowers the energy of a spin up electron 
due to the last term in Eq.~(\ref{eq:Hspin})
and stabilizes the ground state configuration further.
In fact, the second terms of the right-hand side of 
Eq.~(\ref{eq:Hspin})
give rise to an energy gap in the energy spectrum.
As we will show 
using a continuous model for ${\cal H}_{0s}$ in Sec.~\ref{sec:cm}, 
the appearance of the gap will become more clear 
since the term proportional to $\sigma_z$ acts as a mass term
of Dirac fermion.
It is noted that 
$\langle p_n \rangle=0$ is consistent with the presence of a gap,
and non-vanishing $\langle p_m \rangle$ 
gives different signs of the mass terms
for spin up and down electrons.

Next, we consider the next-nearest neighbor (nnn) hopping, $\gamma_n$,
which is an intrinsic perturbation to the edge states.
In the previous paper, we showed that
the nnn interaction, ${\cal H}_{\rm nnn}$,
gives a finite energy bandwidth to the edge states,
$W=\gamma_n$.~\cite{sasaki06apl}
$\gamma_n\approx 0.1\gamma_0$ is obtained 
by first-principles calculation using 
the local density approximation.~\cite{porezag95}
A finite energy band width of the edge states
suppresses the above mentioned 
$\langle p_n\rangle$'s instability 
with respect to a small change of $E_{\rm F}$ 
(see Fig.~\ref{fig:n_pn}).

To see the relationship between 
$\langle p_n \rangle$ and $\langle p_m \rangle$ in detail,
we first define the net pseudo-spin order $P_n$
and the averaged antiferromagnetic order $P_m$:
$P_n \equiv \sum_u |\langle p_n({\bf r}_u) \rangle|/n$ 
and $P_m \equiv \sum_u \langle p_m({\bf r}_u) \rangle/n$
where the summation is taken over
all hexagonal unit cells.
Since not only the edge states but also extended states 
can contribute to $P_i$ ($i=n,m$),
we consider the difference between 
$P_i$ for a tube with the zigzag edges ($P_i^{\rm tube}$)
and that for a corresponding periodic torus system ($P_i^{\rm torus}$)
which does not have edge.
$P_m^{\rm edge}\equiv P_m^{\rm tube}-P_m^{\rm torus}$ 
can be used to show 
the magnetism for the edge states.

%%%%%%%%%%%%%%%%%%%%%%%%%%%%%
\begin{figure}[htbp]
 \begin{center}
  \includegraphics[scale=0.4]{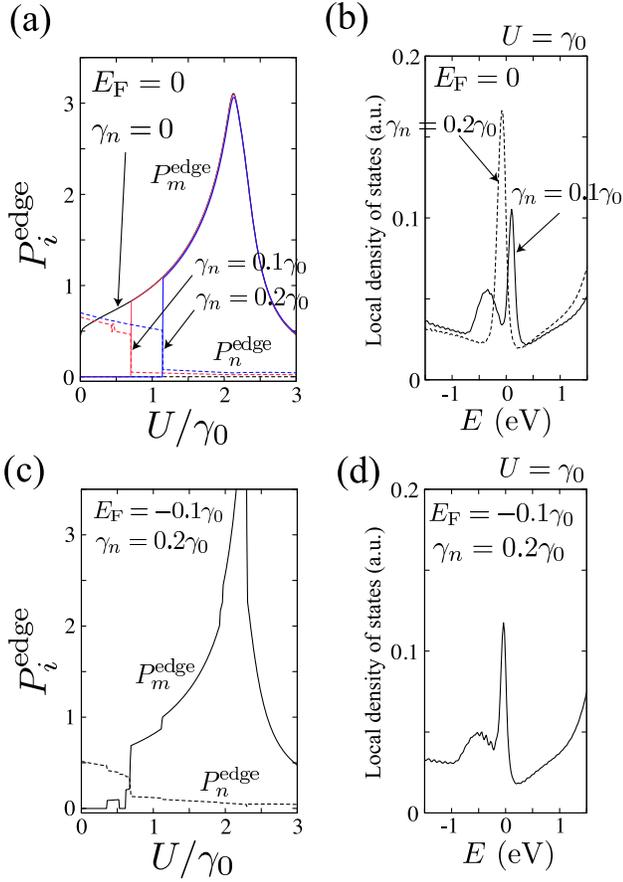}
 \end{center}
 \caption{(color-online) (a)
 Self-consistent solution of 
 $P_m^{\rm edge}$ (solid curve) and $P_n^{\rm edge}$ (dashed curve)
 as a function of $U/\gamma_0$ for 
 $\gamma_n=0$ (black), $0.1\gamma_0$ (red), and $0.2\gamma_0$ (blue)
 in the case of $E_{\rm F}=0$.
 $\gamma_n$ controls the appearance of 
 the pseudo-spin order ($P^{\rm edge}_n$) and of
 the ferrimagnetic order ($P^{\rm edge}_m$).
 (b) The LDOS curves for $\gamma_n=0.1\gamma_0$ (solid curve)
 and $\gamma_n=0.2\gamma_0$ (dashed curve) when $U=\gamma_0$
 and $E_{\rm F}=0$.
 (c) $P_m^{\rm edge}$ (solid curve) and $P_n^{\rm edge}$ (dashed curve)
 for $\gamma_n=0.2\gamma_0$ with $E_{\rm F}=-0.1\gamma_0$.
 This corresponds to that $E_{\rm F}$ is located at the
 center of the edge energy band.
 The small step on $P^{\rm edge}_m$ (at $U/\gamma_0=1$) is 
 due to a finite diameter of a tube. 
 (d) The corresponding LDOS curve when $U = \gamma_0$ in (c).
 The peak structure is sensitive to the position of $E_{\rm F}$.
 }
 \label{fig:pm_int}
\end{figure}
%%%%%%%%%%%%%%%%%%%%%%%%%%%%%

In Fig.~\ref{fig:pm_int}(a), 
we plot $P_m^{\rm edge}$ (solid curve)
and $P_n^{\rm edge}$ (dashed curve)
as a function of $U/\gamma_0$ for 
$\gamma_n=0$ (black), $0.1\gamma_0$ (red), and $0.2\gamma_0$ (blue).
When $\gamma_n=0$, 
no pseudo-spin order
$P^{\rm edge}_n \approx 0$ for any positive value of $U$,
while
the antiferromagnetic order $P^{\rm edge}_m$ 
increases until $U/\gamma_0 \approx 2.1$.
$P^{\rm edge}_m$ decreases when $U/\gamma_0 > 2.1$.
However Fujita {\it et al.} showed that
the magnetism due to the extended states
increases.~\cite{fujita96} 
In case of a finite value of $\gamma_n$, 
the antiferromagnetic order is suppressed 
$P^{\rm edge}_m \approx 0$ up to a finite value of $U$ 
and the magnetism (in the case of $\gamma_n=0$) appears 
discontinuously above the critical value of $U$.
On the other hand, 
the pseudo-spin order appears
$P^{\rm edge}_n \ne 0$ below 
the critical value of $U$.
Thus, when $E_{\rm F}=0$, 
we see that $\gamma_n$ controls 
the occurrence of the pseudo-spin order ($P^{\rm edge}_n$) and
the antiferromagnetic order ($P^{\rm edge}_m$) exclusively.
In Fig.~\ref{fig:pm_int}(b), 
we plot the LDOS curves when $U=\gamma_0$
for $\gamma_n=0.1\gamma_0$ (solid curve)
and $\gamma_n=0.2\gamma_0$ (dashed curve).
When $\gamma_n=0.1\gamma_0$, 
magnetic order is realized so that
there are two peaks in the LDOS curve.
When $\gamma_n=0.2\gamma_0$,
the pseudo-spin order $P^{\rm edge}_n$
is realized 
and there is one peak below the Fermi energy 
in the LDOS curve.
In the case of $U=0$, 
the peak position appears at $E=-\gamma_n$.~\cite{sasaki06apl}
The pseudo-spin order shifts 
the peak position above $E=-\gamma_n$
due to $(U/4) \sum_{{\bf r}_u}\langle p_n \rangle \sigma_z$
in Eq.~(\ref{eq:Hspin}).
This is a possible reason why 
Kobayashi {\it et al.}~\cite{kobayashi05,kobayashi06} 
and Niimi {\it et al.}~\cite{niimi05,niimi06}
observed a peak in the LDOS
below the Fermi energy by 20 $\sim$ 30 meV 
not by $\gamma_n \approx 0.3$ eV.~\cite{porezag95}
Since the most localized edge states have $p_n =\pm 1$
at a unit cell of the edge site 
and the energy is given by $-\gamma_n$,
we see that ${\cal H}_{\rm nnn}=-\gamma_n |p_n|$.
Thus, the $p_n$ dependent energy density 
at a unit cell of the edge site, 
${\cal H}_U+{\cal H}_{\rm nnn}= (U/8) p_n^2- \gamma_n |p_n|+\ldots$,
may become a negative value when the ground state
shows a pseudo-spin order, $\langle p_n \rangle\ne 0$.
In fact, in the case of $\gamma_n=0.2\gamma_0$,
the magnetization disappears even for $U\approx \gamma_0$.

In Fig.~\ref{fig:pm_int}(c), 
we plot $P_m^{\rm edge}$ (solid curve)
and $P_n^{\rm edge}$ (dashed curve)
as a function of $U/\gamma_0$ for 
$\gamma_n = 0.2\gamma_0$ with $E_{\rm F}=-0.1\gamma_0$.
The results show that,
the critical $U$ value decreases 
as compared with the case of $E_{\rm F}=0$
shown by the blue curves in Fig.~\ref{fig:pm_int}(a).
The corresponding LDOS curve for $U = \gamma_0$
is plotted in Fig.~\ref{fig:pm_int}(d).
We see that there are one sharp peak and a broaden peak.
Although $P^{\rm edge}_m \gg P^{\rm edge}_n$
when $U = \gamma_0$,
this two peaks structure is not so clear as the two peaks
in the case of $E_{\rm F}=0$ with $\gamma_n = 0.1\gamma_0$.

\section{Continuous model}\label{sec:cm}

To understand the edge magnetism in detail,
we solve Eq.~(\ref{eq:Hspin}) analytically
by means of a continuous model.
It will be shown that 
the magnetic order is explained 
by a gauge field for the edge states
and spin-dependent mass term.

\begin{figure}[htbp]
 \begin{center}
  \includegraphics[scale=0.4]{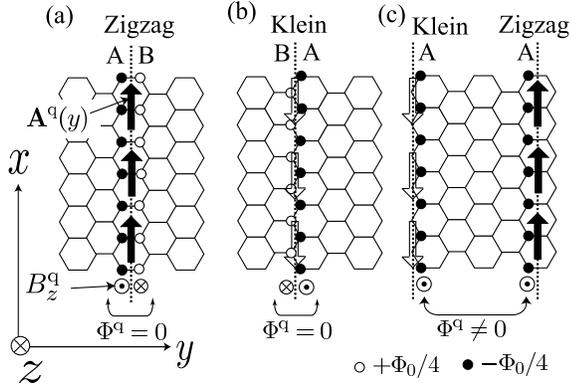}
 \end{center}
 \caption{
 Configuration of ${\bf A}^{\rm q}({\bf r})$ 
 and $B_z^{\rm q}(\mathbf{r})$
 for (a) the zigzag edge and (b) the Klein edge.
 When we cut the graphene sheet at the dashed line,
 the edge and ${\bf A}^{\rm q}({\bf r})=(A_x^{\rm q}(y),0)$ appears.
 The direction of ${\bf A}^{\rm q}({\bf r})$ for the Klein edge 
 is opposite to that of the zigzag edge.
 (c) A graphene with the zigzag edge and the Klein edge.
 }
 \label{fig:gauge}
\end{figure}

The low energy states around $E_{\rm F}=0$ 
consist of electrons near the K-point and K'-point.
Since the K-point and K'-point are related to each other 
by time-reversal symmetry,
it is sufficient to consider only the K-point
when ${\bf B}^{\rm em}=0$.
Then, the low energy Hamiltonian is given by replacing 
${\cal H}_{0s}$ in Eq.~(\ref{eq:Hspin})
with ${\cal H}_{\rm K}$ as
\begin{align}
 \begin{split}
  & {\cal H}_\uparrow
  = {\cal H}_{\rm K}
  + \frac{U}{4} (\langle p_n \rangle - \langle p_m \rangle) \sigma_z
  - \frac{U}{4} \langle m \rangle I,
  \\
  & {\cal H}_\downarrow
  = {\cal H}_{\rm K}
  + \frac{U}{4} (\langle p_n \rangle + \langle p_m \rangle) \sigma_z
  + \frac{U}{4} \langle m \rangle I,
 \end{split}
 \label{eq:Hs_low}
\end{align}
where ${\cal H}_{\rm K}$ is given by
\begin{align}
 {\cal H}_{\rm K} \equiv
 v_{\rm F} \bsigma \cdot (\hat{\bf p} +{\bf A}^{\rm q}({\bf r})).
 \label{eq:H_K}
\end{align}
In Eq.~(\ref{eq:Hs_low}), 
${\cal H}_s$ ($s=\uparrow,\downarrow$)
operates on a two-component wavefunction,
$\psi^{\rm K}_s ={}^t(\psi^{\rm K}_{{\rm A},s},
\psi^{\rm K}_{{\rm B},s})$,
where $\psi_{{\rm A},s}^{\rm K}$ and 
$\psi^{\rm K}_{{\rm B},s}$
are the pseudo-spin up and down states, respectively.
In Eq.~(\ref{eq:H_K}), $v_{\rm F}$ is the Fermi velocity, 
$\hat{\mathbf{p}} \equiv (\hat{p}_x,\hat{p}_y)$ 
is the momentum operator,
and $\bsigma \equiv (\sigma_x,\sigma_y)$ where
$\sigma_i$ ($i=x,y,z$) are the Pauli matrices.
${\bf A}^{\rm q}({\bf r})=(A_x^{\rm q}({\bf r}),A_y^{\rm q}({\bf r}))$ 
is a field that is induced by a defect 
(edge structure)
in the hexagonal unit cell 
and is referred to as the deformation-induced gauge
field.~\cite{kane97,sasaki05} 
The corresponding deformation-induced magnetic field,
$B^{\rm q}_z({\bf r}) \equiv \partial_x A^{\rm q}_y({\bf r}) -
\partial_y A^{\rm q}_x({\bf r})$, 
couples to the pseudo-spin, $\sigma_z$.
This is shown by ${\cal H}_{\rm K}$ squared,
\begin{align}
 {\cal H}_{\rm K}^2 = 
 v_F^2 \left\{ (\hat{\mathbf{p}}+\mathbf{A}^{\rm q}(\mathbf{r}))^2 + \hbar
 B_z^{\rm q}(\mathbf{r}) \sigma_z \right\},
 \label{eq:second-H}
\end{align}
where
$\sigma_z$ selects the direction opposite to 
$B_z^{\rm q}(\mathbf{r})$ in order to decrease the energy.
In the previous paper, we derived ${\bf A}^{\rm q}({\bf r})$
and $B_z^{\rm q}(\mathbf{r})$ for zigzag edges.~\cite{sasaki06jpsj}
${\bf A}^{\rm q}({\bf r})$ appears when we cut 
the hexagonal lattice at dashed lines in Fig.~\ref{fig:gauge}.
Kusakabe {\it et al}. discussed 
the edge state magnetism for the two possible edge
structures,~\cite{kusakabe03} 
that is, the zigzag edge and the Klein edge.~\cite{klein94,klein99}
In Fig.~\ref{fig:gauge}, we show 
the corresponding ${\bf A}^{\rm q}({\bf r})$
and $B_z^{\rm q}(\mathbf{r})$ for 
a zigzag edge (a), a Klein edge (b), and 
a graphene cluster with a zigzag edge at one edge and a Klein edge 
at another edge.
The direction of ${\bf A}^{\rm q}({\bf r})$ for the Klein edge 
is opposite to that of the zigzag edge.
In all cases, we can explain the edge structure
within the same frame.

%%%%%%%%%%%%%%%%%%%%%%%%%%%%%
%%%%%%%%%%%%%%%%%%%%%%%%%%%%%

In the following,
we obtain the wavefunction and the energy eigenvalue $E$
for the following Hamiltonian:
\begin{align}
 \left( {\cal H}_{\rm K} + m_s \sigma_z \right)
 \psi_{p_x,s}^{\rm K}({\mathbf r})=E \psi_{p_x,s}^{\rm K}
 ({\mathbf r}),
 \label{eq:H_low}
\end{align}
where the mass term is given by
\begin{align}
 m_s \equiv  \frac{U}{4} (\langle p_n \rangle \mp \langle p_m \rangle).
 \label{eq:mass}
\end{align}
The mass term depends on real spin, that is, 
the negative (positive) sign 
in front of $\langle p_m \rangle$ 
is for $s=\uparrow$ ($\downarrow$).
Equation~(\ref{eq:H_low})
describes Dirac fermion having a mass, $m_s$,
where the dimension of $m_s$ is energy here.
In obtaining Eq.~(\ref{eq:H_low}), 
we neglect $\mp (U/4) \langle m \rangle$
of Eq.~(\ref{eq:Hs_low}).
The neglected term does not couple to the pseudo-spin 
and only shifts the energy position of each state 
so that it is not important for our discussion.
Further,
we neglect the ${\bf r}$ dependence of $m_s$
in order to simplify the argument.

Since there is translational symmetry along the edge, 
the eigenfunction of Eq.~(\ref{eq:H_low})
can be expressed by
\begin{align}
 \psi_{p_x,s}({\bf r})=N \exp\left(i\frac{p_x}{\hbar} x\right)e^{-G(y)}
 \begin{pmatrix}
  e^{+g_s(y)} \cr e^{-g_s(y)}
 \end{pmatrix},
 \label{eq:wf}
\end{align}
where $s=\uparrow,\downarrow$, and
$x$ ($y$) is parallel (perpendicular) to the zigzag
edge.~\cite{sasaki06jpsj} 
The unknown functions, $G(y)$ and $g_s(y)$, 
and $E$ can be determined by
putting Eq.~(\ref{eq:wf}) into Eq.~(\ref{eq:H_low}).
We obtain
\begin{align}
 & p_x + A_x(y) + \hbar \frac{d}{dy}g_s(y) = D \cosh(2g_s(y)+f_s),
 \label{eq:g} \\
 & \hbar \frac{d}{dy}G(y) = D \sinh(2g_s(y)+f_s),
 \label{eq:G}
\end{align}
where variables $D$ and $f_s$ are defined respectively as
\begin{align}
 & D \equiv \pm \frac{1}{v_F} \sqrt{E^2-m_s^2}, 
 \label{eq:X}
 \\
 & \tanh(f_s) \equiv - \frac{m_s}{E}.
 \label{eq:h}
\end{align}
Here, we consider a localized wavefunction and 
put $G(y)=|y|/\xi$ into Eq.~(\ref{eq:G})
where $\xi$ is localization length of the edge state and 
the zigzag edge is located at $y=0$.
Then we get
\begin{align}
 2g_s(y) + f_s= 
 \begin{cases}
  \displaystyle - \sinh^{-1} \left( \frac{\hbar}{\xi D} \right) & (y < 0) \\
  \displaystyle + \sinh^{-1} \left( \frac{\hbar}{\xi D} \right) & (y > 0).
 \end{cases}
 \label{eq:caseg}
\end{align}
Next, we integrate Eq.~(\ref{eq:g}) with respect to $y$ 
from $-\xi_g$ to $\xi_g$.
By considering $\xi_g \to 0$, 
only singular functions of $A_x^{\rm q}(y)$
and Eq.~(\ref{eq:caseg}) around $y=0$ survive, and we get
\begin{align}
 - \sinh^{-1} \left( \frac{\hbar}{\xi D} \right)
 = \int_{-\xi_g}^{\xi_g} A^{\rm q}_x(y)dy.
 \label{eq:gauge_con}
\end{align}
Using Eqs.~(\ref{eq:caseg}) and (\ref{eq:gauge_con}),
we see from Eq.~(\ref{eq:g}) that 
\begin{align}
 D =
 \frac{p_x}{\cosh\left(\int_{-\xi_g}^{\xi_g} A^{\rm q}_x(y)dy \right)} 
 \label{eq:Xs}
\end{align}
holds except very close to the edge.
From Eqs.~(\ref{eq:gauge_con}) and (\ref{eq:Xs}),
we see that $\xi$ in $G(y)$ is given by 
\begin{align}
 \frac{\hbar}{\xi}
 = -p_x \tanh\left(\int_{-\xi_g}^{\xi_g} A^{\rm q}_x(y)dy \right).
\end{align}
This result is surprising in the sense that
$\xi$ in the presence of $U$ is identical to $\xi$ for
$U=0$.~\cite{sasaki06jpsj}  
The mass term would affect $\xi$, but it is not the case.
The reason for this will be discussed elsewhere.
Finally, we get the energy eigenvalue from
Eqs.~(\ref{eq:X}) and (\ref{eq:Xs}),
\begin{align}
 E^2=m_s^2+
 \frac{(v_Fp_x)^2}{\cosh^2\left(\int_{-\xi_g}^{\xi_g} 
 A^{\rm q}_x(y)dy \right)}.
 \label{eq:Ene}
\end{align}
The energy dispersion relation for the edge states 
of Eq.~(\ref{eq:Ene}) 
is similar to the relativistic energy dispersion relation
for the extended state,
$E^2 = m_s^2 + (v_{\rm F}p_x)^2$.

In Eq.~(\ref{eq:h}), 
we see that 
the sign of $f_s$ depends both on the signs of $m_s$ and $E$.
To obtain a ground state, 
we first consider the valence states $E<0$.
Then, we have $f_s={\rm sign}(m_s)|f_s|$.
Using Eq.~(\ref{eq:gauge_con}), 
we can rewrite Eq.~(\ref{eq:caseg}) as
\begin{align}
 g_s(y) =
 \begin{cases}
  \displaystyle + \frac{1}{2} \int_{-\xi_g}^{\xi_g} 
  A^{\rm q}_x(y')dy' - \frac{1}{2}{\rm sign}(m_s) |f_s| &(y < 0) \\
  \displaystyle - \frac{1}{2} \int_{-\xi_g}^{\xi_g} 
  A^{\rm q}_x(y')dy' - \frac{1}{2}{\rm sign}(m_s) |f_s| &(y > 0).
 \end{cases}
 \label{eq:caseg(y)}
\end{align}
By putting Eq.~(\ref{eq:Ene}) into Eq.~(\ref{eq:h}),
we have 
$|f_s| \approx \left| \int_{-\xi_g}^{\xi_g} A^{\rm q}_x(y)dy \right|$
when $\left| \int_{-\xi_g}^{\xi_g} A^{\rm q}_x(y)dy \right|\gg 0$.
Since $\int_{-\xi_g}^{\xi_g} A^{\rm q}_x(y)dy \gg 0$ 
for the zigzag edge (see Fig.~\ref{fig:gauge}(a)), 
we get from Eq.~(\ref{eq:caseg(y)}) 
\begin{align}
 g_\uparrow(y) \approx 
 \begin{cases}
  \displaystyle \int_{-\xi_g}^{\xi_g} A^{\rm q}_x(y')dy'  &(y < 0) \\
  \displaystyle 0 \ &(y > 0),
 \end{cases}
 \label{eq:gup}
\end{align}
when $m_\uparrow < 0$.
The wavefunction of this spin up state appears 
near the edge consisting of {\rm A}-atoms ($y<0$) 
in the valence band ($E<0$):
$\psi_{p_x,\uparrow}(y<0) \propto \exp(-|y|/\xi){}^t(1,0)$.
The wavefunction at the edge consisting of {\rm B}-atoms ($y>0$)
is pseudo-spin unpolarized and
the amplitude is negligible 
due to the normalization of the wavefunction.
Similarly, for $m_\downarrow > 0$, we have
\begin{align}
 g_\downarrow(y) \approx 
 \begin{cases}
  \displaystyle 0 &(y < 0) \\
  \displaystyle -\int_{-\xi_g}^{\xi_g} A^{\rm q}_x(y')dy' &(y > 0).
 \end{cases}
 \label{eq:gdn}
\end{align}
The corresponding wavefunction of spin down state
is pseudo-spin down state 
appearing only near the edge consisted of {\rm B}-atoms ($y>0$).
It is noted that the spin for a conduction edge state ($E>0$)
is opposite to that of a valence edge state.
Thus, we obtain local ferrimagnetism near the zigzag edge.

\begin{figure}[htbp]
 \begin{center}
  \includegraphics[scale=0.4]{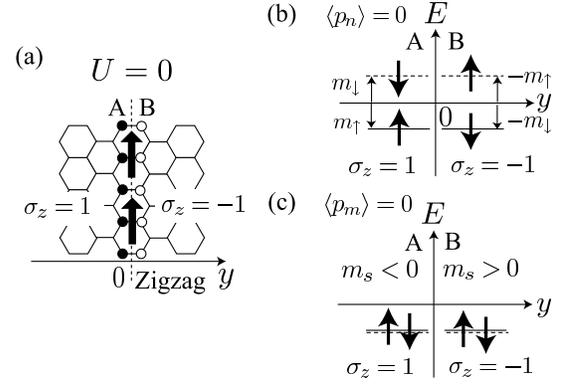}
 \end{center}
 \caption{
 (a)
 Deformation-induced gauge (magnetic) field produces 
 asymmetry of pseudo-spin.
 The pseudo-spin polarized localized state (i.e., the edge states)
 appears in pair with respect to $y=0$.
 (b)
 Due to the mass term, this symmetric character with respect to $y=0$ 
 is lost and the edge state can appear independently
 at $y > 0$ and $y<0$
 (spatial parity with respect to $y\to -y$ is broken
 by the mass term).
 When the sign of the mass is opposite with respect to spin, 
 namely when $m_\uparrow=-m_\downarrow$,
 spin becomes asymmetric with respect to $E=0$, 
 and a local ferrimagnetism appears.
 (c) In case that $m_\uparrow=m_\downarrow$,
 spin up and down states are degenerate and 
 both appear below the Fermi level 
 when $m_s <0$ ($m_s >0$) for $y<0$ ($y>0$)
 so that the ground state is spin unpolarized.
 }
 \label{fig:spectrum}
\end{figure}

In Fig.~\ref{fig:spectrum},
we show how the magnetism appears around the zigzag edge.
In Fig.~\ref{fig:spectrum}(a),
if $U=0$, 
the opposite direction of pseudo-spin (the edge state) appears
both for $y>0$ ($\sigma_z=-1$) and $y<0$ ($\sigma_z=1$).
However when $U\ne 0$, due to the mass term
the edge state exists only one of the two sides 
in Fig.~\ref{fig:spectrum}(a).
When pseudo-spin order is suppressed 
($\langle p_n \rangle =0$) and
antiferromagnetic order appears 
($\langle p_m \rangle \ne0$),
we get a situation 
of a different sign for $m_\uparrow$ and $m_\downarrow$
($m_\uparrow=-m_\downarrow$) from Eq.~(\ref{eq:mass}).
In this case, 
up (down) spin edge state appears for $y<0$ ($y>0$) 
for the valence band
according to Eqs.~(\ref{eq:gup}) and (\ref{eq:gdn}), 
while down (up) spin edge state appears
for $y<0$ ($y>0$) for the conduction band.
It is consistent with the numerical result of 
Fujita {\it et al.} for $E_{\rm F}=0$,~\cite{fujita96}
in which the electrons are occupied only for the valence band.
If $E_{\rm F}$ is shifted from $E_{\rm F}=0$ then
$\langle p_m \rangle$ will disappear.

When pseudo-spin order appears 
($\langle p_n \rangle \ne 0$) and
antiferromagnetic order is suppressed
($\langle p_m \rangle =0$),
we get another situation that
the sign of $m_\uparrow$ and $m_\downarrow$ are the same
($m_\uparrow=m_\downarrow$) from Eq.~(\ref{eq:mass}).
In this case,
the energy levels for up and down spin edge states
are degenerate even for $U\ne 0$
(see Fig.~\ref{fig:spectrum}(c)).
Thus the ground state is not spin polarized 
but pseudo-spin polarized.
Even in case that $m_\uparrow \ne m_\downarrow$,
spin up and down edge states both appear
below the Fermi level so that the ground state is still 
spin unpolarized
if ${\rm sign}(m_\uparrow)={\rm sign}(m_\downarrow)$.
It is interesting to note that, in Fig.~\ref{fig:spectrum}(c),
the energy level position
for $y<0$ can appear above (below) $E_{\rm F}=0$ 
when $m_s > 0$ ($m_s < 0$).
The sign of $m_s$, that is, 
$m_s <0$ ($m_s >0$) for $y<0$ and $m_s >0$ ($m_s <0$) for $y>0$,
is consistent with the numerical results
given in Sec.~\ref{sec:ana}

%%%%%%%%%%%%%%%%%%%%%%%%%%%%%
%%%%%%%%%%%%%%%%%%%%%%%%%%%%%

As we have shown in Sec.~\ref{sec:ana},
the magnetism of the edge states is affected by 
the nnn hopping that can stabilize $\langle p_n \rangle$.
In the continuous model,
we showed in the previous paper 
that the nnn perturbation works as 
\begin{align}
 {\cal H}_{\rm nnn}
 = \gamma_n \frac{\ell^2}{\hbar} B_z^{\rm q}({\bf r})\sigma_z,
 \label{eq:nnn}
\end{align}
for the edge states where 
$\ell \equiv 3a_{\rm cc}/2$.~\cite{sasaki06jpsj} 
${\cal H}_{\rm nnn}$ is proportional to $\sigma_z$ so that
${\cal H}_{\rm nnn}$ appears as an additional term for the mass.
If $\gamma_n$ is sufficiently large then
we have $m_\uparrow = m_\downarrow$ and magnetism disappears.
This is consistent with the numerical results given in
Sec.~\ref{sec:ana}.

\section{Discussion}\label{sec:dis}

A magnetism of the edge states 
would give rise to two LDOS peaks 
since only spin up (or down) edge states are located below
the Fermi energy to give a spin polarization in the ground state.
Although we have examined this mechanism using the Hubbard model,
the appearance of two peaks 
seems to be a model independent consequence of the magnetism 
of the edge states.
The LDOS near the zigzag edge of graphite
has been measured by
STS,~\cite{klusek00,kobayashi05,kobayashi06,niimi05,niimi06} 
but no experimental group has observed the two peaks
in the STS data.
It is possible that the position of the Fermi energy 
in these experiments
is not suitable for the occurrence of the magnetism 
(see Fig.~\ref{fig:pm_int}).
Thus if we change of the Fermi energy,
LDOS will give a split of the peak,
which is an evidence that the edge states form a magnetism.

We explained the magnetism of the edge states 
in terms of the spin dependent mass terms and 
the deformation induced gauge field.
It is known that 
the mass and a gauge field in the Weyl equation
induce the parity anomaly
in the ground state.~\cite{semenoff84,PhysRevD.29.2375,redlich84}
The mass term in Eq.~(\ref{eq:H_low})
changes its sign under spatial parity with respect to $y \to -y$
and $\psi_{p_x,s}^{\rm K}\to \sigma_x \psi_{p_x,s}^{\rm K}$.
The mass term violates the spatial parity and
can induce a quantum anomaly in the ground state,
which is referred to as the parity anomaly. 
By applying the formula of the parity anomaly~\cite{semenoff84} 
to our case, 
we obtain 
\begin{align}
 \langle \rho_s({\bf r}) \rangle = 
 \frac{1}{2} \frac{B^{\rm q}_z({\bf r})}{\Phi_0} {\rm sign}(m_s) 
 + \ldots,
\end{align}
where $\Phi_0=2\pi \hbar$ is the flux quantum
and correction may arise due to 
higher order derivatives of $B^{\rm q}_z({\bf r})$.
In the case of $m_\uparrow= -m_\downarrow$,
we have magnetism, i.e.,
$\langle \rho_\uparrow({\bf r})-\rho_\downarrow({\bf r}) \rangle \ne 0$.
Moreover,
using $\int_{\rm unit \ cell} B^{\rm q}_z({\bf r})d^2{\bf r}=\pm
\Phi_0/4$ that will be derived in the following,
the magnetization at the edge is estimated by 
$\langle m \rangle = \pm 1/4$, which 
is good agreement with our numerical result, 
$\langle m \rangle_{0,L} \approx \pm 0.25$,
shown in Fig.~\ref{fig:density_U}.
Thus, we think that there is a close relationship between 
the edge states magnetism and the parity anomaly.
In fact, the anomaly survives 
even in the massless limit $m_s \to 0$,
which is consistent to the fact that 
an infinitesimal value of $U$ gives a finite magnetism
if we do not consider the nnn interaction
(see 
$P_m^{\rm edge}$ in Fig.~\ref{fig:pm_int}(a) 
and Fig.~5 in Ref.~\citen{fujita96}).
A graphene with the zigzag edge and the Klein edge
can be used to know that the deformation induced magnetic field
in a unit cell at one side of the edge is given by 
$\int_{\rm unit \ cell} B^{\rm q}_z({\bf r})d^2{\bf r}=\pm
\Phi_0/4$ (see Fig.~\ref{fig:gauge}(c)).
In this case, 
$\Phi^{\rm q}=\int B_z^{\rm q}({\bf r})d^2{\bf r}$ is nonzero and
the index theorem~\cite{bertlmann00}
can be used to know $B_z^{\rm q}({\bf r})$.
The theorem states that ${\cal H}_{\rm K}$ possesses
$|\Phi^{\rm q}/\Phi_0|$ zero energy edge states.
Since it is known that the number of the zero energy states 
in a $(n,0)$ nanotube is given by $n$,~\cite{sasaki05prb} 
then we can know that 
the flux in a unit cell at the zigzag edge,
$\Phi_u$, is given by $\pm \Phi_0/4$.
Here, we used $2 \times (2n \Phi_u) = \pm n \Phi_0$.
The factor 2 comes from the time-reversal symmetry (the K and K' points)
and the factor $2n$ is the total number of edge sites at the zigzag
edge and the Klein edge.
$|\Phi_u|=\Phi_0/4$ 
is consistent with the numerical result 
by Nakada {\it et al}.~\cite{nakada96} 
who demonstrate that an edge shape with 
three or four zigzag sites 
per sequence is sufficient to show an edge state.

In summary,
we have shown that 
the instability of the pseudo-spin order of the edge states
induces ferrimagnetic order 
in the presence of the Coulomb interaction.
The nnn hopping can stabilize the pseudo-spin order, but
a larger value of $U$ makes the pseudo-spin order unpolarized 
and gives rise to a ferrimagnetic order.
The ferrimagnetic order is sensitive to the Fermi energy position.
In case that the pseudo-spin order is realized
one peak appears in the LDOS near the zigzag edge,
which is consistent to the existing experimental results.
Using a continuous model of the Weyl equation,
we showed that the deformation-induced gauge field
gives rise to the magnetism of the edge states
if the mass terms have different sign for different spin 
edge states.

\section*{Acknowledgments}

 Authors would like to thank M. Suzuki and K. Nomura 
 for fruitful discussion.
 R. S. acknowledges a Grant-in-Aid (No. 16076201) from MEXT.
 
%\bibliographystyle{apsrev}
%\bibliographystyle{prsty2}
% \bibliography{../../../../bib/sasaki}

\end{document}